\documentclass[twoside,a4paper,11pt]{sea10}
\usepackage{graphicx}
\usepackage{hyperref}
\usepackage{movie15}
\newcommand {\apgt} {\ {\raise-.5ex\hbox{$\buildrel>\over\sim$}}\ }
\newcommand {\aplt} {\ {\raise-.5ex\hbox{$\buildrel<\over\sim$}}\ }
\begin{document}
\pagenumbering{arabic}
\pagestyle{myheadings}
\thispagestyle{empty}
{\flushleft\includegraphics[width=\textwidth,bb=58 650 590 680]{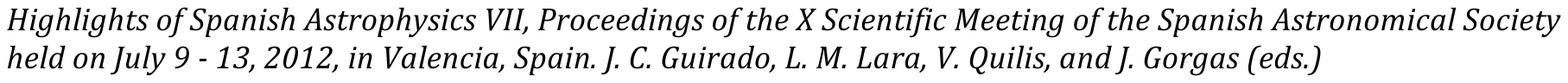}}
\vspace*{0.2cm}
\begin{flushleft}
{\bf {\LARGE
%
Discerning the location of the $\gamma$-ray emission region in blazars from multi-messenger observations
%
}\\
\vspace*{1cm}
%
Iv\'{a}n Agudo$^{1,2}$,
Alan P. Marscher$^{2}$, 
Svetlana G. Jorstad$^{2,3}$, 
and 
Jos\'{e} L. G\'{o}mez$^{1}$
%
}\\
\vspace*{0.5cm}
%
$^{1}$
{Instituto de Astrof\'{i}sica de Andaluc\'{i}a, CSIC, Apartado 3004, 18080, Granada, Spain}\\
$^{2}$
{Institute for Astrophysical Research, Boston University, 725 Commonwealth Avenue, Boston, MA 02215, USA}\\
$^{3}$
{Astronomical Institute, St. Petersburg State University, Universitetskij Pr. 28, Petrodvorets, 198504 St. Petersburg, Russia}
%
\end{flushleft}
%
\markboth{
The location of the $\gamma$-ray emission region in blazars from multi-messenger observations.
}{ 
%
Agudo et al.
%
}
\thispagestyle{empty}
\vspace*{0.4cm}
\begin{minipage}[l]{0.09\textwidth}
\ 
\end{minipage}
\begin{minipage}[r]{0.9\textwidth}
\vspace{1cm}
\section*{Abstract}{\small
%
Relativistic jets in AGN in general, and in blazars in particular, are the most energetic and among the most powerful astrophysical objects known so far. 
Their relativistic nature provides them with the ability to emit profusely at all spectral ranges from radio wavelengths to $\gamma$-rays, as well as to vary extremely at time scales from hours to years. 
Since the birth of $\gamma$-ray astronomy, locating the origin of $\gamma$-ray emission has been a fundamental problem for the knowledge of the emission processes involved. 
Deep and densely time sampled monitoring programs with the \emph{Fermi} Gamma-ray Space Telescope and other facilities at most of the available spectral ranges (including millimeter interferometric imaging and polarization measurements wherever possible) are starting to shed light for the case of blazars. 
After a short review of the status of the problem, we summarize two of our latest results --obtained from the comprehensive monitoring data compiled by the Boston University Blazar monitoring program -- that locate the GeV flaring emission of the BL Lac objects AO~0235+164 and OJ287 within the jets of these blazars, at $\apgt12$ parsecs from the central AGN engine.
%
\normalsize}
\end{minipage}
%
%
%
\section{Introduction: Blazars}
The most luminous long-lived sources of radiation in the cosmos -- active galactic nuclei (AGN) -- are powered by gas falling from an accretion disk into a super-massive black hole (SMBH, $\sim10^{8}$ times more massive than the Sun) at their centre \cite{LyndenBell:1969p14815,Shakura:1973p14818}. 
The most exotic AGN are blazars, a class defined by wild variability of flux of non-thermal radiation from radio to $\gamma$-ray energies. 
Members of this class include BL Lacertae objects (BL Lacs), and flat spectrum radio quasars (FSRQ, the high power version of the former). 
The remarkable properties of blazars include superluminal apparent motions as high as $40\,c$ or more \cite{Jorstad:2005p264}, substantial changes in flux and linear polarization on time scales as short as hours, and extremely variable $\gamma$-ray luminosities that can exceed those at other wavebands by as much as 3 orders of magnitude. 
These phenomena are caused by relativistic jets of highly energized, magnetized plasma that are propelled along the rotational poles of the SMBH-disk system \cite{Blandford:1982p6283}. 
AGN jets pointing within $<10^{\circ}$ of our line of sight beam their radiation and shorten the variability time scales to give blazars their extreme properties.

\section{The problem of the location of $\gamma$-ray emission in blazars}

\begin{figure}
\center
\includegraphics[width=10cm]{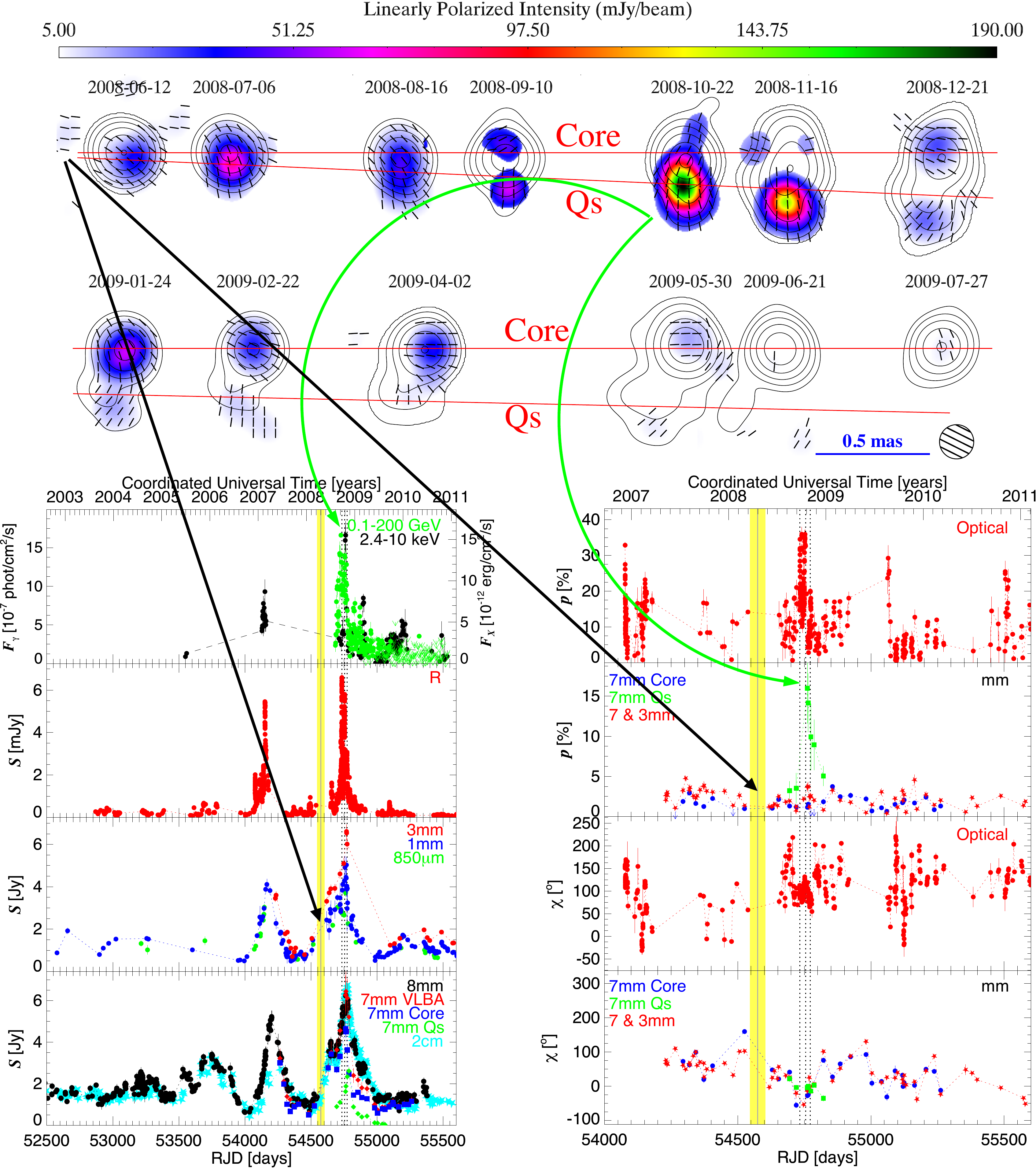}
\caption{\label{fig1} \footnotesize{{\it Top:} Sequences of 0.15 mas resolution VLBA images of AO~0235+164 at 7\,mm. Contour levels represent the total intensity map, whereas the color scale and the superimposed line segments symbolize the polarized intensity and the orientation of the electric vector position angle, respectively. {\it Left:} Light curves of AO~0235+164 at $\gamma$-rays, X-rays, optical and mm-wavelengths (from top to bottom). The vertical line denotes the time of ejection of superluminal component Qs in the relativistic jet of AO~0235+164. {\it Right:} Optical and mm-wave linear polarization of AO~0235+164 as a function of time. Reproduced from \cite{Agudo:2011p15946}.}}
\end{figure}

Explaining the location of the $\gamma$-ray emitting regions in blazar jets remains one of the greatest current challenges in high-energy astrophysics. 
Having a good knowledge of such location is essential to shed new light on other fundamental questions such as: what is the origin of the seed photons responsible for such $\gamma$-ray emission? what is the emission mechanism producing the high energy emission? or what is the extragalactic relativistic jet composition at different distances from the SMBH (electromagnetic, $\rm{e}^-\rm{e}^+$, $\rm{e}^-\rm{p}^+$) and how does it change between them?

There are essentially two main locations of the $\gamma$-ray emission in blazar jets that are usually claimed in the literature \cite[e.g.]{Dotson:2012p17671}.
One of them -- the \emph{near site}, from $\apgt0.1$ to $\aplt1$\,pc from the central engine -- has been the scenario considered until recently.
This region, inside the broad line region of AGN (BLR), is embedded by the BLR's optical--UV photon field that can be scattered to $\gamma$-rays by relativistic electrons in the jet plasma.
This inverse Compton emission process can be treated by existing multi--spectral--range emission models.
The near site has been frequently invoked to explain the short time-scales of $\gamma$-ray variability of a few hours (or less) reported in some blazars \cite[e.g.]{Tavecchio:2010p14858}.
This is however not always a good argument, because causality arguments only imply that short time scales are related to small emission regions, not that such emission regions should be at any particular location.
Another argument in favor of the $\gamma$-ray emission being produced in the near site region is that if the $\gamma$-ray emission is produced within the BLR, it is possible to explain the sharp breaks at a few GeV seen in the $\gamma$-ray spectra of \emph{some} blazars by opacity to pair production by (hydrogen and helium) emission lines in the broad line region \cite[e.g.]{Poutanen:2010p12551}.

However, there is an increasing number of blazars for which the $\gamma$-ray and radio-mm jet emission sites -- the latter at several parsecs far from the SMBH \cite[e.g.]{Marscher:2008p15675}, where the jet starts to be visible at millimeter wavelengths with VLBI-- seem to be co-spatial \cite{Marscher:2010p11374,Jorstad:2010p11830,Agudo:2011p14707,Agudo:2011p15946}.
This is the \emph{far site} scenario, where the radiation field from the BLR is not relevant
This scenario is also invoqued to explain, the correlation found between the $\gamma$-ray flux and the radio and mm flux in large blazar samples \cite[e.g.]{Pushkarev:2010p17672,LeonTavares:2011p16196}.

\section{Our observing approach to solve the problem}

Since the physical conditions of blazar jets are so extreme, theoretical progress in understanding them depends on guidance from comprehensive observational programs that study blazars in a variety of ways. 
The \emph{Fermi} Gamma-ray Space Telescope (\emph{Fermi} hereafter), the new $\gamma$-ray observatory launched by NASA in mid 2008, represents a revolution in time coverage, energy range, sensitivity, and angular resolution with regard to its predecessor, the Compton Gamma-Ray Observatory (CGRO). 

\begin{figure}
\center
\includegraphics[width=13cm]{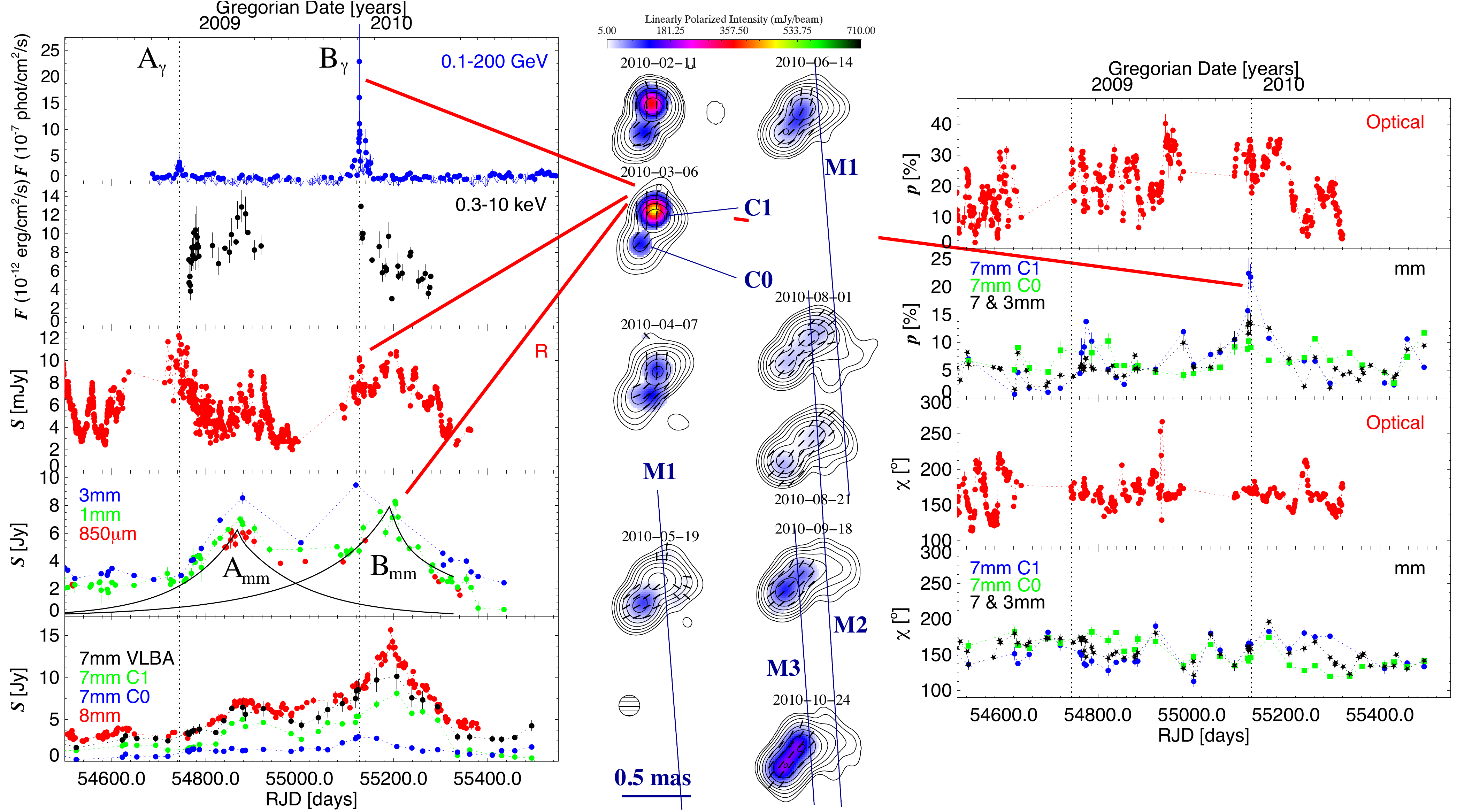}
\caption{\label{fig2} \footnotesize{{\it Left:} Light curves of OJ287 at $\gamma$-rays, X-rays, optical and mm-wavelengths (from top to bottom). The vertical lines denote the times of peak $\gamma$-ray flux of flares ${\rm{A}}_{\gamma}$ and ${\rm{B}}_{\gamma}$. 
{\it Center:} Sequences of 0.15 mas resolution VLBA images of OJ287 at 7\,mm from Feb. to Oct. 2010. Contour levels represent the total intensity map, whereas the color scale and the superimposed line segments symbolize the polarized intensity and the orientation of the EVPA, respectively.
{\it Right:} Optical and mm-wave linear polarization of OJ287 as a function of time. Reproduced from \cite{Agudo:2011p14707}}}
\end{figure}

Since 2008, several international consortia have been devoting their efforts to support \emph{Fermi} observations at all other accessible spectral ranges. 
Among such monitoring programs the Boston University Blazar monitoring program routinely observes 34 bright
$\gamma$-ray blazars with monthly time sampling at the highest possible angular resolution at mm observing wavelengths, with the Very Long Baseline Array (VLBA, in the USA), and with Global Millimetre VLBI Array (GMVA, composed of 5 European mm stations plus the VLBA). 
Millimeter VLBI optimally images the innermost regions of jets in blazars where mm emission is not strongly affected by synchrotron opacity, nor Faraday rotation of the polarized emission. 
Moreover, this data set is complemented, within the collaboration, by single dish radio and mm observations (including the IRAM 30m Telescope), and optical polarimetric observations (with telescopes at the Lowell, Calar Alto, St. Petersburg State University, Steward, and Crimean Observatories), plus a number of other non polarimetric telescopes, with time samplings from a few to many observations per month for each object. 
These are complemented with past RXTE X-ray observations (time sampling $<1$ week) of some of the brightest sources and with daily sampled $\gamma$-ray light curves by the \emph{Fermi} observatory, which (past and future) data became public from August 2009. 
Less frequently, observations with the X-ray, UV, and optical Swift satellite are also performed. 
Through correlations and measurement of time lags, such a program allows the establishment of connections - and therefore relative locations - of the emission at different wavebands, and of associations of the variable emission with bright knots that move down the jet, usually at superluminal apparent velocities. 
Furthermore, the polarization reveals the degree of ordering and mean direction of the magnetic field, and provides further means for evaluating simultaneous events at different wavebands.

\section{New results}

\subsection{$\gamma$-ray Flaring Emission in the Parsec-Scale Jet of Blazar AO~0235+164}

In \cite{Agudo:2011p15946}, we present the time coincidence in the end of 2008 of the propagation of the brightest superluminal feature detected in AO~0235+164 (Qs) with an extreme multi-spectral-range ($\gamma$-ray to radio) outburst, and an extremely high optical and 7\,mm (for Qs) polarization degree, which provides strong evidence supporting that all these events are related (Fig.\ref{fig1}). 
This is confirmed at high statistical significance by probability arguments and Monte-Carlo simulations \cite{Agudo:2011p15946}. 
These simulations show the unambiguous correlation of the $\gamma$-ray flaring state in the end of 2008 with those in the optical, millimeter, and radio regime, as well as the connection of a prominent X-ray flare in October 2008, and of a series of optical linear polarization peaks, with the set of events in the end of 2008.
The observations are interpreted as the propagation of an extended moving perturbation through a re-collimation structure at the end of the jet's acceleration and collimation zone, and locate the $\gamma$-ray and lower frequency emission in flares of the BL~Lac object AO~0235+164 at $\apgt12$\,pc in the jet of the source from the central engine.
 
\subsection{$\gamma$-ray Flaring Emission in Blazar OJ287 $\apgt14$\,pc from the Black Hole}

In \cite{Agudo:2011p14707}, the correlation between the brightest $\gamma$-ray and mm flares is also found to be statistically significant.
The two $\gamma$-ray peaks, detected by {\emph Fermi}-LAT, that we report in that paper happened at the rising phase of two exceptionally bright mm flares accompanied by sharp linear polarization peaks (see also Fig.~\ref{fig2}).
The VLBA images show that these mm flares in total flux and polarization degree occurred in a jet region at $>14$\,pc from the innermost jet region.
The time coincidence of the brighter  $\gamma$-ray flare and its corresponding mm linear polarization peak evidences that both the $\gamma$-ray and mm outbursts occur $>14$\,pc from the central black hole.
We find two sharp optical flares occurring at the peak times of the two reported $\gamma$-ray flares.
All this is interpreted as the $\gamma$-ray flares being produced by synchrotron self-Compton scattering of optical photons from the flares triggered by the interaction of moving knots with a stationary conical shock in the jet. 

\section{Conclusions}

We have presented the results obtained from the BL Lac objects OJ~287 and AO~0235+164 through multi-spectral range monitoring data from the BU Blazar monitoring program.
In both cases, the analysis of these comprehensive data sets support --without inferences from any kind of modeling, but just from the interpretation of the observing data-- that the high energy $\gamma$-ray flaring emission is produced in the relativistic jets associated to these sources and at large distances ($>12$\,pc) from the corresponding central SMBHs, hence supporting the \emph{far site} scenario for the generation of $\gamma$-ray emission in flares.

This evidence rejects the possibility of production of $\gamma$-ray emission in flares as produced by seed photons from the broad line region in OJ~287 and AO~0235+164, which BL Lac type nature also prevents from displaying prominent emission from the broad line region.
Moreover, the fact that both OJ~287 and AO~0235+164 are classified as BL Lac objects -- for which no IR emission from the dusty torus surrounding the AGN has been detected so far-- disfavors an scenario where the $\gamma$-ray emission is produced by the inverse Compton process by upscatering of the dusty-torus photon field.
In \cite{Agudo:2011p15946}, we provide further arguments against this latter scenario for AO~0235+164 (but see \cite{Ackermann:2012p751159}).

Therefore, both for OJ~287 and AO~0235+164, we favor a mechanism for the generation of prominent $\gamma$-ray emission in flares where the seed low energy photons upscatered to $\gamma$-rays come from the synchrotron emission radiated from the jet's plasma, which is a third natural source of optical--IR photons. 
We note though that there are in the literature other cases for which this conclusion is not as clear as for the two cases presented here \cite[e.g.]{Abdo:2010a, Abdo:2010b,Abdo:2011,Vercellone:2011,Hayashida:2012}.
Hence, it seems that our results cannot be generalized for the entire blazar class.
%
%
\small  
%
\section*{Acknowledgments}   
The authors acknowledge funding support from MICIIN grant AYA2010-14844; CEIC grant P09-FQM-4784; NASA grants NNX08AJ64G, NNX08AU02G, NNX08AV61G, and NNX08AV65G; NSF grant AST-0907893; and NRAO award GSSP07-0009.
The VLBA is an instrument of the NRAO, a facility of the NSF operated under cooperative agreement by AUI. 
The PRISM camera at Lowell Observatory was developed by Janes et al. 
The Calar Alto Observatory is jointly operated by MPIA and IAA-CSIC. 
The IRAM 30\,m Telescope is supported by INSU/CNRS, MPG, and IGN. 

\begin{thebibliography}{}
\small
%
\bibitem{LyndenBell:1969p14815}{Lynden-Bell, D. 1969, Nature, 223, 690}
\bibitem{Shakura:1973p14818}{Shakura, N. I. \& Sunyaev, R. A. 1973, A\&A, 24, 337}
\bibitem{Jorstad:2005p264}{Jorstad, S. G. et al. 2005, AJ, 130, 1418}
\bibitem{Blandford:1982p6283}{Blandford, R. D. \& Payne D. G. 1982, MNRAS, 199, 883}
\bibitem{Dotson:2012p17671}{Dotson, A. et al. 2012, ApJ, 758, L15}
\bibitem{Tavecchio:2010p14858}{Tavecchio, F. et al. 2010, MNRAS, 405, L94}
\bibitem{Poutanen:2010p12551}{Poutanen, J. \& Stern, B. 2010, ApJ, 717, L118}
\bibitem{Marscher:2008p15675} {Marscher, A. P. et~al. 2008, Nature , 452, 966}
\bibitem{Marscher:2010p11374} {Marscher, A. P. et~al. 2010, ApJ, 710, L126}
\bibitem{Jorstad:2010p11830} {Jorstad, S. G. et~al. 2010, ApJ, 715, 362}
\bibitem{Agudo:2011p14707} {Agudo, I. et~al. 2011a, ApJ, 726, L13}
\bibitem{Agudo:2011p15946} {Agudo, I. et~al. 2011b, ApJ, 735, L10}
\bibitem{Pushkarev:2010p17672}{Pushkarev, A. B., Kovalev, Y. Y. \& Lister, M. L. 2010, ApJ, 722, L17}
\bibitem{LeonTavares:2011p16196}{Le\'{o}n-Tavares, J. et al. 2011, A\&A, 532, A146}
\bibitem{Ackermann:2012p751159}{Ackermann, M. et al. 2012, A\&A, 751, 159}
\bibitem{Abdo:2010a}{Abdo, A. A. et al. 2010a, ApJ, 721, 1425}
\bibitem{Abdo:2010b}{Abdo, A. A. et al. 2010b, Nature, 463, 191}
\bibitem{Abdo:2011}{Abdo, A. A. et al. 2011, ApJ, 726, 43}
\bibitem{Vercellone:2011}{Vercellone, S. et al. 2011, ApJ, 736, L38}
\bibitem{Hayashida:2012}{Hayashida, M. et al. 2012, ApJ, 754, 114}
%
\end{thebibliography}
\end{document}